\documentclass[prb,preprint]{revtex4-1} 

\usepackage{amsmath}  
\usepackage{amsfonts} 
\usepackage{graphicx} 

\usepackage{graphicx}
\usepackage{epic}
\usepackage{eepic}
\usepackage{latexsym}
\usepackage{amssymb}

\newcommand{\eq}[1]{(\ref{#1})}
\newcommand{\be}{\begin{equation}}
\newcommand{\ee}{\end{equation}}
\newcommand{\bea}{\begin{eqnarray}}
\newcommand{\eea}{\end{eqnarray}}

\newcommand{\vs}[1]{\vspace{#1 mm}}
\newcommand{\hs}[1]{\hspace{#1 mm}}

\def\C{\Gamma}
\def\d{\delta}
\def\D{\Delta}
\def\e{\epsilon}

\def\f{\phi}
\def\fr{\frac}

\def\l{\lambda}
\def\L{\Lambda}
\def\m{\mu}
\def\n{\nu}
\def\p{\pi}

\def\r{\rho}
\def\s{\sigma}

\def\O{\Omega}

\def\o{\omega}

\def\del{\partial}

\let\bm=\bibitem
\def\nn{\nonumber}

\newcommand{\phys}{\textrm{phys}}

\begin{document}

\title{The Cosmological Constant Problem: An Accessible  Introduction}

\author{Ali Kaya}

\email[]{alikaya@tamu.edu}

\author{Adam Lahey}

\email[]{adamlahey03@tamu.edu}

\affiliation{\vs{3}Department of Physics and Astronomy, Texas A\&M University, College Station, TX 77843, USA \vs{10}}

\begin{abstract}
\vs{5}

We present a pedagogical introduction to the cosmological constant problem that requires only basic knowledge of quantum field theory and general relativity. A massive real scalar field is used to illustrate how the quantum vacuum energy density and pressure can be calculated both in flat spacetime and in an expanding universe. Detailed computations are provided for dimensional, cutoff, and adiabatic regularizations. No attempt is made to address quantum gravitational effects, and the expanding-universe background is treated classically. We point out that although the commonly cited discrepancy of 120 orders of magnitude between theory and observation is based on an estimate that does not account for regularization and renormalization, fundamental principles of quantum field theory nevertheless lead to a huge mismatch. In addition to this large discrepancy, we emphasize that there are also conceptual challenges related to cosmic expansion, such as the choice between comoving and physical scales in certain contexts and the non-uniqueness of vacuum.

\end{abstract}

\maketitle

\section{Introduction} 

In physics, a clash between well-established theories and observations often indicates a fundamental lack of understanding. A successful resolution of such a conflict usually leads to a significant breakthrough and a major paradigm shift involving new physics. One of the best-known historical examples is the explanation of blackbody radiation. Combining the basic principles of statistical physics with classical electromagnetic theory leads to the Rayleigh–Jeans law, which gives the correct energy distribution at long wavelengths but fails dramatically at short wavelengths. It yields  a divergent result, a problem historically known as the ultraviolet catastrophe. Planck’s resolution of this problem,which introduced the revolutionary concept of energy quantization, was the first fundamental step toward the development of quantum mechanics.

There is a similar clash today which involves quantum physics and general relativity: the well-known cosmological constant problem \cite{wein1} (see Ref. \onlinecite{cch} for a historical account). The ground state of a quantum harmonic oscillator has a positive energy. At the same time, a free bosonic quantum field, such as the Higgs, contains infinitely many degrees of freedom and can be viewed as a collection of infinitely many harmonic oscillators. Consequently, the ground state energy (density) in quantum field theory diverges, much like the Rayleigh–Jeans ultraviolet catastrophe in blackbody radiation. 

There are established techniques in field theory, such as regularization and renormalization, that allow us to handle these infinities and obtain finite, physically meaningful results. These methods have been tested with great success in countless particle-collider experiments. Meanwhile, the Friedmann equation, which is derived from applying general relativity to a cosmological background, states that any form of energy density contributes to the Hubble parameter, the expansion rate of the universe. The problem is that the observed total energy density of the universe, inferred from the present-day Hubble constant, differs enormously from the theoretical prediction for the vacuum (ground state) energy density of a quantum field such as the Higgs.

In this paper, we aim to provide an accessible, semi-technical introduction to the cosmological constant problem, similar to \cite{ajp} (see Refs. \onlinecite{r1,r2} for more technical reviews). In addition to deriving the well-known large discrepancy in the vacuum energy density in flat spacetime, an aspect typically emphasized in discussions of the cosmological constant problem, we also address issues arising from cosmic expansion. This perspective is necessary for consistency but is often overlooked.

By definition, the cosmological constant has a specific equation of state:  its pressure is equal to minus its energy density. We will discuss how the pressure associated with the ground state of a scalar quantum field can be determined using basic thermodynamic arguments, both in flat spacetime and in an expanding universe. This is again a crucial aspect that is often neglected in this context. We provide explicit computational details for dimensional, cutoff, and adiabatic regularization methods, although a full discussion of renormalization theory lies beyond the scope of this paper. As we will see, cosmic expansion introduces complications for the standard regularization techniques that work consistently in flat spacetime.

Finally, we will highlight deeper conceptual difficulties related to the cosmological constant problem. In particular, we elaborate on the non-uniqueness of the vacuum in an expanding universe. In some instances, there is an ambiguity in choosing between comoving and physical scales, and as we will discuss, one can find arguments supporting either choice. Our aim is to emphasize that the difficulty is not merely technical but it has a deep conceptual origin at the intersection of quantum theory, gravity, and cosmology. Hence its resolution may require a profound revision of our understanding of these fundamental subjects.  

\section{Dimensions, Units and the Friedmann Equation} 

General relativity starts from the assumption that spacetime is a four-dimensional pseudo-Riemannian manifold endowed with a Lorentzian metric, having one timelike and three spacelike directions. One can always introduce local coordinates $(t, x^i)$, $i=1,2,3$ so that $t$ and $x^i$ parameterize timelike and spacelike directions, respectively. It is natural to assign the dimension of time  [$\mathcal{T}$] to $t$ and the dimension of length [$\mathcal{L}$] to $x^i$. It is convenient to introduce $x^\m=(ct,x^i)$, $\m=0,1,2,3$, so that all components of  $x^\m$ have the dimension [$\mathcal{L}$]. 

The metric defines the infinitesimal invariant distance square $ds^2$ between two nearby points
\be
ds^2=g_{\m\n}dx^\m dx^\n.
\ee
The distance square $ds^2$, which has dimension [$\mathcal{L}^2$], can be positive, negative, or zero, corresponding to spacelike, timelike, and null-separated points, respectively. The metric $g_{\m\n}$ is assumed to be dimensionless. 

According to general relativity, the phenomenon of gravity is associated with the spacetime curvature, which is determined by the distribution of matter (subject to certain assumptions, such as boundary conditions). Matter is characterized by the stress-energy-momentum tensor $T_{\m\n}$ whose components, when referred to a suitable basis, become energy density, momentum density, pressure and stress. Thus, unlike in Newtonian gravity, mass/energy density is not the sole source of gravity. The metric tensor itself intrinsically enters the definition of $T_{\m\n}$, for instance, in the definition of the spatial volume required for calculating energy density.  The dimension of $T_{\m\n}$ is that of energy density,  [$\mathcal{M/(LT}^2)$], where  [$\mathcal{M}$] denotes mass dimension. 

The specific space-time curvature determined by $T_{\m\n}$ is the Ricci tensor $R_{\m\n}$, which  is calculated straightforwardly from the derivatives of $g_{\m\n}$ and its inverse $g^{\m\n}$. It is obtained from the Riemann curvature tensor $R^{\mu}{}_{\n\r\s}$ by a contraction: $R_{\m\n}=R^{\l}{}_{\m\l\n}$.  Both the Riemann curvature tensor and the Ricci tensor involve exactly two derivatives with respect to the coordinates $x^\m$; consequently, the dimension of $R_{\m\n}$ is  [$\mathcal{L}^{-2}$]. The Ricci scalar $R$ is defined as $R=g^{\m\n}R_{\m\n}$. The Riemann tensor is a direct measure of intrinsic curvature, vanishing if and only if the region is flat. Since the Ricci tensor is obtained from the Riemann tensor by a contraction, it measures a sense of averaged-out curvature. 

The Einstein field equations are written as:
\be\label{ae} 
R_{\mu\nu}-\fr12 g_{\m\n}R=\fr{8\pi G}{c^4} T_{\mu\nu},
\ee
These equations constitute a system of non-linear second-order partial differential equations for the metric tensor components $g_{\m\n}$, provided the matter content, that is, the information on the right-hand side,  is specified. The precise combination of the fundamental constants on the right-hand side, $8\pi G/c^4$, is determined by the requirement that in the limit of weak gravity and non-relativistic speeds, the equations recover Newtonian gravity. In that limit, the Newtonian potential can be identified as a perturbation around the flat-spacetime metric $\eta_{\m\n}$ and all components of the stress-energy-momentum tensor $T_{\m\n}$ become negligible except for the energy density component. 

The stress-energy-momentum tensor must satisfy the conservation equation
\be\label{cons1} 
\nabla_\m T^{\m\n}=0,
\ee
where $\nabla_\m$ is the covariant derivative of the metric $g_{\m\n}$. For timelike and spatial values of the index $\n$, \eq{cons1} ensures the conservation of energy and momentum, respectively. This is consistent with the right-hand side of \eq{ae}, where $\nabla^\m\,(R_{\mu\nu}-\fr12 g_{\m\n}R)=0$ is satisfied  identically, which is known as  the contracted Bianchi identity. 

In the context of general relativity, it is convenient to introduce geometrized units so that all physical quantities acquire the dimension of length (see e.g. Ref. \onlinecite{wald}). However, when quantum effects are also incorporated, converting all quantities to mass (or energy) dimension is often much more useful. This offers a direct way to compare gravitational effects with quantum phenomena arising in particle physics.

First, any time dimension $[\mathcal{T}]$  can be converted to length dimension $[\mathcal{L}]$ using the speed of light $c$ (we have already applied this conversion above by defining the time coordinate as $ct$). Then, using the constant combination 
 $c/\hbar$, which has dimension $[1/\mathcal{(ML)}]$, any length dimension can be converted to mass dimension $[\mathcal{M}]$.
For instance, one can define coordinates $\tilde{x}^\m=c x^\m/\hbar$ which will have mass dimension -1, i.e.  $[1/\mathcal{M}]$. Similarly, defining the scaled Ricci tensor and stress-energy-momentum tensor as
\bea
&&\tilde{R}_{\m\n}=\fr{\hbar^2}{c^2}R_{\m\n},\\
&&\tilde{T}_{\m\n}=\fr{\hbar^3}{c^5}T_{\m\n}
\eea
results in a Ricci tensor with mass dimensions 2 and a stress-energy-momentum tensor with mass dimension 4.  In terms of these new variables, the Einstein field equations become
\be\label{eam} 
\tilde{R}_{\mu\nu}-\fr12 g_{\m\n}\tilde{R}=\fr{8\pi }{M_p^2} \tilde{T}_{\mu\nu},
\ee
where $M_p=\sqrt{c\hbar/G}$ is the Planck mass. The dimensional consistency of \eq{eam} is now manifest. 

In units where $c=1$ and $\hbar=1$, the numerical conversion factors between the original and the scaled quantities become unity.  We will adopt this convention and drop the tilde from the scaled variables, typically working with mass-dimension quantities (the conversion back to the original set is straightforward). Although the preceding dimensional analysis is elementary, it will be very useful in acknowledging the severity of the cosmological constant problem.

For a spatially flat cosmological model, the metric is assumed to have the following form
\be\label{frw} 
ds^2=-dt^2+a(t)^2(dx^2+dy^2+dz^2),
\ee
where $a(t)$ is known as the scale factor of the universe. This metric respects the observed homogeneity and isotropy of the universe on cosmological scales. Consistency with these symmetries also requires that the stress-energy-momentum tensor $T_{\m\n}$ has only components for energy density and pressure, taking the diagonal form 
\be \label{diag} 
T_{\m\n}=\textrm{diag}(\r,P,P,P),
\ee
where $\r=\r(t)$ and $P=P(t)$. 

Generally, there exists a relation between the energy density $\rho$ and the pressure $P$ that follows from the physical properties of matter. This relation is called the equation of state, and for almost all cosmologically relevant forms of matter, it takes the form
\be
P= w\,\r,
\ee
where the constant $w$ is called the equation of state parameter. For example, non-relativistic matter (dust) with negligible peculiar velocities has no pressure, corresponding to $w = 0$. Electromagnetic radiation has $w = 1/3$, while a cosmological constant is, by definition, characterized by $w = -1$. If more than one type of matter is present, the total stress-energy tensor is simply the sum of the individual contributions.

With the above assumptions, Einstein’s field equations reduce to two nonlinear equations for $a(t)$, involving its first and second order time derivatives. Deriving these equations is somewhat involved, as it requires computing the Christoffel symbols $\C^{\m}{}_{\n\r}$, the Riemann curvature tensor $R^{\m}{}_{\n\r\s}$, and finally the Ricci tensor $R_{\m\n}$. The $(tt)$ or equivalently the $(00)$ component of \eq{ae}, which is first order in time derivatives, is known as the Friedmann equation
\be\label{fr} 
H^2=\fr{8\pi }{3M_p^2} \r,
\ee
where 
\be
H=a^{-1} \, \fr{da}{dt}
\ee
is the Hubble parameter characterizing the expansion rate of the universe. The fact that the energy density on the right-hand side is suppressed by the square of the Planck mass $M_p$ is the main reason why the cosmic expansion rate is small and difficult to observe.

Within a certain accuracy, the currently observed value of the Hubble parameter is given by $H_0\simeq70\, $km/s$\cdot$Mpc, where Mpc (megaparsec) is approximately $3\times 10^{19}$ km. It can also be stated as  $H_0\simeq 2.3\times 10^{-18}\,{\textrm s}^{-1}$. To better appreciate the smallness of this number, one can express it in units of energy. Since $H$ is given in  $[1/\mathcal{T}]$,  $H/c$ has dimension $[1/\mathcal{L}]$, $\hbar H/c^2$ has the dimension of mass $[\mathcal{M}]$  and $\hbar H$ has the dimension of energy. Using the value of $\hbar$ one finds $H_0\simeq 1.5 \times 10^{-33}\,$ eV (electron volts), which corresponds to an extremely small energy scale. For comparison, the electron mass is about $0.5\,$ MeV (mega-electron volts). 

Using $M_p\,c^2\simeq 1.22\times 10^{19}\, $GeV (giga-electron volts), the current energy density $\r_0$ can be estimated from \eq{fr}  as
\be\label{r0}
\r_0\simeq 4 \times 10^{-11}\,(\textrm{eV})^4,
\ee
which accounts for  all components including dust, radiation, dark matter and dark energy. 

\section{Vacuum Energy Density and the Cosmological Constant Problem} 

A key insight from quantum mechanics is that, in contrast to classical physics, the vacuum exhibits nontrivial structure. For instance, the vacuum state of a quantum harmonic oscillator has a so-called zero-point energy of $\fr12 \hbar \o$, where $\o$ is the oscillator frequency.  In a sense, this can be attributed to the uncertainty principle, which, loosely speaking, prohibits a state of complete rest. Not surprisingly, the vacuum in quantum field theory exhibits an even richer structure.

In this section, with the Higgs field in mind, we determine the vacuum energy density of a free real scalar field $\f$ with mass $m$ in flat spacetime (the expanding universe case will be addressed in subsequent sections). Like all other components, this energy density contributes to the right-hand side of the Friedmann equation \eq{fr}, and thus to the expansion rate of the universe. In general, such a calculation requires quantizing the scalar field, defining the corresponding energy density operator, and evaluating its vacuum expectation value. However, a free quantum scalar field is equivalent to an infinite collection of harmonic oscillators, one for each Fourier mode. That is, when the field is Fourier-decomposed in momentum space, the modes decouple, and each becomes a harmonic oscillator (see e.g. Ref. \onlinecite{ps}). 

It is best to describe the decoupling  when the spatial directions are compact circles. In that case, we can introduce coordinates $(x,y,z)$ and assume that  $x\equiv x+L$, $y\equiv y+L$ and $z\equiv z+L$, where $L$ is the size of each compact direction.  For the Fourier modes $\exp[i(k_x x+ k_y y+ k_z z)]$ to become single-valued, the momentum components must be quantized as
\be\label{kc} 
 k_x=\fr{2\pi}{L}\,n_x, \hs{5} k_y=\fr{2\pi}{L}\,n_y,\hs{5}k_z=\fr{2\pi}{L}\,n_z, 
 \ee
where $(n_x,n_y,n_z)$ are integers.  

After performing a Fourier transformation, the field in real space $\f(x,y,z,t)$ is replaced by its momentum-space counterpart $\tilde{\f}(k_x,k_y,k_z,t)$. The canonical quantization then implies that, {\it for each} $\vec{k}=(k_x,k_y,k_z)$, the mode $\tilde{\f}$ behaves  as a simple harmonic oscillator with frequency $\o(\vec{k})=(k_x^2+k_y^2+k_z^2+m^2)^{1/2}$ (in units  $c=1$ and $\hbar=1$). The Hilbert space for each oscillator can be constructed as usual by repeatedly acting with the creation operator $a^\dagger_{\vec{k}}$  on the ground state $(a^\dagger_{\vec{k}})^n\left|0\right>$, where $n\geq0$ is an integer. The total Hilbert space of the quantum field is the tensor product of all these individual oscillator Hilbert spaces, which is known as the Fock space. For a given oscillator labeled by $\vec{k}$, the state  $(a^\dagger_{\vec{k}})^n\left|0\right>$ has the energy $E_n=(n+\fr12)\,\o(\vec{k})$,  and it represents  $n$ particles each carrying momentum $\vec{k}$ and energy $\o(\vec{k})$, with $n=0$ being the ground state.

The total ground-state energy of the quantum field is the sum of the ground-state energies of all the oscillators
\be\label{kce} 
E_0=\sum_{\vec{k}}\fr12 \o(\vec{k})=\sum_{k_x,k_y,k_z}\fr12\sqrt{k^2+m^2},
\ee
where $k^2=k_x^2+k_y^2+k_z^2$ and the components of $\vec{k}$ are given in \eq{kc}. It is possible to convert this sum into an integral in the limit $L\to\infty$. Defining 
\be
\D k_x=\D k_y=\D k_z=\fr{2\pi}{L},
\ee
where $2\pi/L$ is the basic spacing in momentum space, one can write \eq{kce} as
\be
E_0=\fr{L^3}{(2\pi)^3}\sum_{k_x,k_y,k_z}\fr12\sqrt{k^2+m^2}\,\D k_x \D k_y \D k_z.
\ee
Then, in the limit $L\to\infty$, the sum can be replaced by an integral so that
\be\label{r1} 
\rho=\fr{1}{2(2\pi)^3}\int_{-\infty}^{+\infty} d^3 k\, \sqrt{k^2+m^2},
\ee
where the vacuum energy density $\rho$ is defined as
\be\label{el} 
\rho=\fr{E_0}{L^3}.
\ee
Notably, the final expression for the vacuum energy density,  \eq{r1},  is independent of  $L$. 

The integral in \eq{r1} is unfortunately divergent. Such infinities arise in almost all physical quantities calculated in field theory, primarily because a field generally represents infinitely many degrees of freedom (in our case, the free scalar field becomes an infinite sum of simple harmonic oscillators). There is a systematic procedure for dealing with these infinities in quantum field theory, known as regularization and renormalization.

Regularization is an intermediate step in which the divergent integral is rendered finite through some mechanism. For example, instead of integrating momenta in \eq{r1} from $-\infty$ to $+\infty$, one can introduce a cutoff $\L$ and restrict the integration to the region$(-\L,+\L)$, so that the integral remains finite as long as $\L$ is finite. This simple method is called cutoff regularization. 

Another widely used technique is to define the integral in \eq{r1} in $d$-dimensions rather then three. In general, the integral can be expressed in such a way that the result depends analytically on $d$ and remains finite except at certain simple poles. This approach is called dimensional regularization.

Ultimately, one wishes to remove the regularization by taking the limits $\L\to\infty$ or $d\to3$ in the cutoff and dimensional regularization schemes, respectively. The resulting expression typically contains both divergent and finite parts. The systematic procedure of subtracting the divergent contributions through the introduction of counterterms and of redefining the physical parameters, when possible, is known as renormalization.

Discussing renormalization theory is beyond the scope of this work; however, it is relatively straightforward and quite instructive to apply dimensional regularization to \eq{r1}. To do so, one can rewrite the integral in $d$-dimensions as
\be\label{r2} 
\rho_d=\fr{\m^{3-d}}{2(2\pi)^d}\int_{-\infty}^{+\infty} d^d k\, \sqrt{k^2+m^2}=\fr{\m^{3-d}}{2(2\pi)^d}\int_{0}^{\infty} dk\,\O_{d-1} k^{d-1} \sqrt{k^2+m^2}
\ee
so that \eq{r1} corresponds to the case $d=3$. Here, a constant mass scale $\m$ is introduced to ensure that the mass dimension of $\r_d$ remains  4 (and as we will see, introducing $\m$ renders  the expressions dimensionally consistent in the limit  $d\to3$).  In the last expression in \eq{r2}, we switched to spherical coordinates in momentum space where $k$ is the radial coordinate. The angular integration yields $\O_{d-1}$, the surface area of the unit $(d-1)$-dimensional sphere with the values $\O_1=2\pi$, $\O_2=4\pi$ and in general  
\be
\O_{d-1}=\fr{2\pi^{d/2}}{\C(\fr{d}{2})},
\ee
where $\C$ denotes the gamma function. 

The integral in \eq{r2} remains divergent for any real value of $d$. However, by extending $d$ to the complex plane, the integral can be evaluated in terms of Gamma functions (see e.g. Ref. \onlinecite{ps}, p. 249-250), yielding
\bea
\r_d&=&\fr{\m^{3-d}}{2(2\pi)^d}\O_{d-1}\left[-\fr{m^{d+1}}{4\sqrt{\pi}}\C\left(\fr{d}{2}\right)\C\left(-\fr{d+1}{2}\right)\right]\nn\\
\nn\\
&=&-\fr{\m^{3-d}m^{d+1}}{2^{d+2}\pi^{(d+1)/2}}\,\C\left(-\fr{d+1}{2}\right). \label{r3} 
\eea
The Gamma function is analytic in the complex plane, except for simple poles at non-positive integers $n=0,-1,-2,-3...$ and near $z=-n$ it behaves as 
\be\label{g} 
\C(z)=\fr{(-1)^n}{n!(z+n)}+\textrm{finite terms}.
\ee
Defining $\e=3-d$, one can expand \eq{r3} in powers of $\e$ near $\e=0$ obtaining (see e.g. Ref. \onlinecite{ps}, p. 250)
\be\label{r4}
\rho=-\fr{m^4}{32\pi^2}\fr{1}{\e}+\fr{m^4}{32\pi^2}\left[\ln(m/\m)+C\right]+{\cal O}(\e^2),
\ee
where the numerical constant $C$ comes from the finite terms in \eq{g} as well as from the powers of 2 and $\pi$. This final expression highlights the importance of introducing the parameter $\m$ in \eq{r2}; it ensures that the argument of the logarithm is dimensionless.

The isolated infinity in \eq{r4} can be removed, yielding the finite result 
\be\label{r5}
\rho=\fr{m^4}{32\pi^2}\left[\ln(m/\m)+C\right].
\ee
This removal is justified within renormalization theory through a cancellation mechanism involving counterterms.
When the divergent part is subtracted, a finite arbitrariness remains, characterized by the presence of the arbitrary renormalization scale $\m$ in \eq{r5}. These ambiguities are resolved, again within renormalization theory, by imposing a finite number of renormalization conditions at the scale $\mu$, which in turn define the measurable physical parameters of the theory, such as masses and coupling constants.

As a result, the standard reasoning of quantum field theory, which has been successfully applied to explain a vast range of physical phenomena, implies that the vacuum energy density $\r$ associated with a free massive scalar field must be proportional to $m^4$. Using the experimentally measured Higgs mass value $m=125\,$ GeV one obtains an energy density of the order of 
\be\label{r6} 
\r\sim 10^{36}\, (\textrm{eV})^4. 
\ee
There is, therefore,  an enormous discrepancy between the energy density \eqref{r0} inferred from the expansion of the universe and the value \eqref{r6} predicted by field theory.

Recall that the cosmological constant is characterized by the equation of state $P=-\r$  and it is possible to see that the quantum vacuum satisfies this property. The pressure $P$ can be determined from the first law of thermodynamics
\be\label{fl} 
dE=-PdV,
\ee
where $E$ and $V$ are the energy and the volume of the system, respectively. No heat exchange is assumed, $dQ=0$, as there is no environment or other system to interact with. The energy-volume relation $E(V)$ follows from \eq{el}, giving $E=\r V$, where $V=L^3$ and $\r$ does not depend on $V$ . Substituting this relation into the first law, \eqref{fl} immediately yields the cosmological constant equation of state,
\be
P=-\r.
\ee
This result could also have been anticipated from Lorentz symmetry, which requires the vacuum energy–momentum tensor to take the form $T_{\m\n}\propto \eta_{\m\n}$, where $\eta_{\m\n}$ is the flat Minkowski metric. Although regularization procedures can sometimes break symmetries, dimensional regularization is known to largely preserve them.
 
Extra care is required when applying dimensional regularization to massless fields, since infrared (IR) divergences typically appear as $k\to0$ in momentum space integrals. For a massless scalar field, another convenient choice is the cutoff regularization and from \eq{r1} one  finds
\be\label{lfs} 
\r=\fr{1}{16\pi^2}\L^4.
\ee
The cutoff scale $\L$ is usually associated with a scale above which a theory breaks down. For the Standard Model, this is expected to be well above the GeV scale, hence still leading to a large mismatch with \eq{el}. 

Naively setting $\L=M_{pl}$ in \eq{lfs} yields the often-quoted 120 orders-of-magnitude difference with \eq{r0}. However, in this case, as in dimensional regularization, a subtraction must be performed, and the result should be interpreted within the framework of renormalization theory. Therefore, the figure of 120 should be treated with caution.

\section{Incorporating Cosmic Expansion} 

In the previous section, we saw that the vacuum energy density of a massive scalar field, calculated using well-established techniques, yields a result that is incompatible with the expansion of the universe as dictated by the Friedmann equation. However, this calculation was performed in flat spacetime, without taking into account the cosmic expansion, an approach that is inherently inconsistent. Since the vacuum energy drives the expansion, which in turn can alter the vacuum energy density, both must be treated self-consistently within a unified framework. Incorporating the effects of cosmic expansion introduces additional complications and renders the cosmological constant problem even more intriguing.

As pointed out above, a {\it spatially flat} cosmological spacetime can be described by the metric \eq{frw}. One can still assume the spatial coordinates to be periodic with period $L$. However, $L$ is now referred to a {\it coordinate} or  {\it comoving} length scale.  A fundamental precept of general relativity is that coordinates have no direct physical meaning; they can only be interpreted through the metric tensor.

If one takes a displacement four-vector, say $\D x^\m=(0,\D x,0,0)$, its squared length can be calculated using the metric tensor as  $g_{\m\n}\D x^\m \D x^\n=a(t)^2 \D x^2$ (in principle, only infinitesimal coordinate intervals must be considered, but a finite spatial separation is well defined here since the constant-time slices are flat). Therefore, one can define a physical length $\D x_{\phys}$ corresponding to a spatial coordinate difference $\D x$, such that 
\be
\D x_{\phys}=a(t) \D x. 
\ee
Similarly, the momentum variable $\vec{k}$,  as it appears in the Fourier decomposition, represents a comoving vector (in  four-vector notation, it can be viewed as the spatial components of a covariant vector $k_\m$).  Then, for a given comoving scale $k=(k_x^2+k_y^2+k_x^2)^{1/2}$, there corresponds a physical momentum scale $k_{\phys}$ given by 
\be
k_{\phys}=\fr{k}{a(t)}.
\ee
One can also define comoving and physical wavelengths  $\l=1/k$ and $\l_{\phys}=1/k_{\phys}=a(t)\l$. For a fixed comoving wavelength $\l$, $\l_{\phys}$ increases for increasing $a(t)$, illustrating how the wavelength is stretched by the cosmic expansion. 

\section{Slow Adiabatic Expansion } \label{V} 

To study the effects of cosmic expansion on the vacuum energy density, we first assume that the expansion is {\it very slow}  and that the modes evolve {\it adiabatically.}  We will later assess the validity of this assumption and show that it is, in fact, not a good approximation throughout cosmic history. Nevertheless, it is instructive to examine the complications that arise even in the adiabatic case.

In an expanding universe described by the metric \eq{frw}, the scalar field can still be Fourier decomposed in the spatially flat space. In the free theory,  the modes, labeled by a comoving wave vector $\vec{k}$, again decouple from one another, and their quantization is equivalent to that of a simple harmonic oscillator. When the expansion is slow and the modes evolve adiabatically their energy, to a very good approximation, is given by (in what follows, the time argument of $a(t)$ will not be shown explicitly, in order to simplify the notation)
\be\label{wad} 
\o(\vec{k})=\sqrt{k_{\phys}^2+m^2}=\sqrt{\fr{k^2}{a^2}+m^2}. 
\ee
The energy density $\r$ is obtained by dividing the ground-state energy by the {\it physical volume,}
 \be\label{re} 
\rho=\fr{E_0}{a^3L^3}.
\ee 
Following the same steps as after \eq{kce}, it is straightforward to obtain
\be\label{re1} 
\rho=\fr{1}{2(2\p a)^3}\int_{-\infty}^{+\infty} d^3 k\, \sqrt{\fr{k^2}{a^2}+m^2}. 
\ee
It is crucial to note the scale factor $a(t)$ dependence of this last expression. Naively, changing the integration variable $k\to a k$ removes all $a(t)$ dependence and reduces \eq{re1} to the flat spacetime expression \eq{r1}. However, this substitution cannot be justified since \eq{re1} is divergent and thus ill defined. 

To proceed, dimensional regularization can be performed in essentially the same way. The energy density $\r_d$ in $d$-spatial dimensions can be written as 
\be\label{re2} 
\rho_d=\fr{\m^{3-d}}{2(2\pi a)^d}\int_{-\infty}^{+\infty} d^d k\, \sqrt{\fr{k^2}{a^2}+m^2}=\fr{(\m/a)^{3-d}}{2(2\pi)^d}\int_{0}^{\infty} dk_{\phys}\,\O_{d-1} k_{\phys}^{d-1} \sqrt{k_{\phys}^2+m^2}. 
\ee
Note that the overall dependence on the scale factor changes from $1/a^3$ to  $1/a^d$, since the $d$-dimensional spatial volume scales as $a^d$. In the last line, the substitution $k_{\phys}=k/a$ is justified because the integral is convergent and well defined for complex values of $d$. Taking the $d\to3$ limit and removing the simple pole infinity as before\cite{ft1} yields 
\be\label{r5e}
\rho=\fr{m^4}{32\pi^2}\left[\ln(ma/\m)+C\right].
\ee
Although the result is of the same order of magnitude as in flat spacetime, the possible appearance of the $\ln(a)$ dependence in such quantities has generated significant debate in the literature (see e.g. Refs. \onlinecite{ln1,ln2}). 

To grasp one of the main reasons behind the debate, let us first note that $\mu$ is a comoving scale, as it is introduced to compensate for the dimensional mismatch associated with $\vec{k}$ defined in $d$-dimensions. One then has two options: either treat $\mu$ as a constant, which leaves the $\ln(a)$ dependence, or assume that the physical scale $\mu_{\text{\phys}}$ is constant, which implies $\mu = a\,\mu_{\text{\phys}}$. In the latter case, the scale-factor dependence disappears from the logarithm, yielding $\ln(m/\mu_{\phys})$. 

There is a similar ambiguity in the cutoff regularization of a massless scalar field. By setting $m=0$ in \eq{re1} and introducing a comoving cutoff $\L$ one obtains
\be\label{cutoffr}
\r=\fr{1}{16\pi^2}\fr{\L^4}{a^4}=\fr{1}{16\pi^2}\L^4_{\phys}
\ee
Once again, one may choose to treat either $\L$ or the physical cutoff $\L_{\phys}=\L/a$ as a fixed scale, which respectively preserves or removes the $a(t)$-dependence in \eq{cutoffr}.

It may seem more reasonable to keep the physical scale fixed since, as mentioned above, the variables associated with the coordinates do not have direct physical meanings. However, if $\L$ varies with time, the number of degrees of freedom in the system changes accordingly, because the modes are labeled by the comoving vector $\vec{k}$, which is bounded by $k<\L$. A decreasing $\L$ implies that some degrees of freedom are lost, while an increasing $\L$ leads to the emergence of new ones. This, in turn, causes energy-momentum non-conservation, which is a major issue. 

To elaborate on this point,  one can rewrite the first law of thermodynamics \eq{fl} in terms of $\r$ by using $E=V \r$ and $V=a^3L^3$, which gives 
\be\label{econs}
\fr{d\r}{dt}+3H(\r+P)=0. 
\ee
This is the general statement of energy conservation in the cosmological setting, following directly from the conservation equation \eq{cons1} for the energy-momentum tensor \eq{diag} in the background metric \eq{frw}. 

For the massless scalar field, setting $m=0$ in \eq{wad}, and using \eq{kc} and \eq{kce} shows that the ground state energy scales as $E_0\propto 1/(aL)=1/V^{1/3}$.  The thermodynamic relation $P=-\del E/\del V$, which follows from the first law \eq{fl}, then yields the equation of state
\be\label{es1} 
P=\fr13 \r.
\ee
Integrating \eq{econs} with \eq{es1} gives $\rho \propto a^{-4}$. By comparing this result with \eq{cutoffr}, one finds that energy conservation requires $\L$ to remain constant, rather than keeping the physical cutoff $\L_{\phys}$ constant.

Since the equation of state \eq{es1} corresponds to radiation, the vacuum energy density of a massless scalar field with cutoff regularization appears to imply a {\it cosmological radiation problem} rather than a cosmological constant problem. This distinction is crucial, as it leads to drastically different consequences for cosmic evolution.

To appreciate the difference between these scenarios, we first note that for all physically viable matter, the equation of state parameter falls in the range
\be\label{esr} 
-1\leq w \leq1. 
\ee
This follows from the dominant energy condition, which requires a non-negative energy density and forbids energy flow faster than light (see e.g. Ref. \onlinecite{car}). One can see from \eq{econs} that for any component with an equation of state parameter $w$ satisfying \eq{esr}, the energy density $\rho$ decreases with expansion, except in the special case of the cosmological constant ($w= -1$), for which $\rho$ remains constant. Therefore, the cosmological constant is expected to eventually dominate cosmic expansion, rendering all other components negligible.\cite{wald2} 

Even with the assumption that the modes evolve adiabatically, it remains difficult to determine the ultimate impact of the vacuum energy density on cosmic evolution, specifically, whether it behaves like radiation ($\rho \propto a^{-4}$) or a cosmological constant ($\rho =$ constant), which leads to quite different possibilities.

\section{General Case} 

The above analysis provides useful intuition, but it relies crucially on the assumption of very slow cosmic expansion, an assumption that is clearly invalid in the early universe. We now turn to the general case, in which no such approximation is imposed. 

In a curved background, the equation of motion for a minimally coupled massive scalar field takes the form
\be\label{nn} 
\left(g^{\m\n}\nabla_\m\nabla_\n-m^2\right)\f=0.
\ee
In the spatially flat cosmological metric \eq{frw}, Fourier decomposition of the field leads to the mode equation
\be\label{me} 
\fr{d^2 \f_k}{dt^2}+3H\fr{d\f_k}{dt}+\left(\fr{k^2}{a^2}+m^2\right)\f_k=0,
\ee
where $\f_k=\f_k(t)$ and $k$ is a comoving momentum label. 

In the classical theory, \eq{me} describes the evolution of a wave with comoving wave number $k$ in an expanding universe. In the quantum theory, the mode functions are used to introduce the creation and annihilation operators $a_{\vec{k}}$ and $a_{\vec{k}}^\dagger$ so that the field operator $\f$ takes the form
\be\label{mex} 
\f=\fr{1}{(2\pi)^{3/2}}\int d^3k \left[\f_k(t)e^{i\vec{k}.\vec{r}}a_{\vec{k}}+\f_k^*(t)e^{-i\vec{k}.\vec{r}}a^\dagger_{\vec{k}}\right],
\ee
where $\vec{k}.\vec{r}=k_x x+ k_y y+ k_z z$. 
As in the quantum harmonic oscillator, the vacuum of the theory is defined by
\be\label{vac} 
a_{\vec{k}}\left|0\right>=0,
\ee
for all $\vec{k}$.  

From the action yielding the equations of motion \eq{nn}, the momentum $P_\f$ conjugate to $\f$ in a cosmological spacetime can be found as $P_\f=a^3(\del\f/\del t)$. The equal time commutation relation between $\f$ and $P_\f$, 
which is the central requirement of canonical quantization, is satisfied if one imposes $[a_{\vec{k}},a^\dagger_{\vec{k'}}]=\d^3(\vec{k}-\vec{k'})$ and the  Wronskian condition
\be\label{wc} 
\f_k \fr{d\f_k^*}{dt}-\f_k^* \fr{d\f_k}{dt}=\fr{i}{a^3},
\ee
where $\d^3(\vec{k}-\vec{k'})$ is the three-dimensional Dirac delta function. The Hilbert space with a positive-definite inner product is then constructed by acting with creation operators on the vacuum, $a^\dagger_{\vec{k}_1}a^\dagger_{\vec{k}_2}...a^\dagger_{\vec{k}_n}\left|0\right>$, 
which represents an $n$-particle state carrying comoving momenta $\vec{k}_1$, $\vec{k}_2$, ... , $\vec{k}_n$, one for each particle. 

Without any additional requirement, the canonical quantization procedure is completed, and the theory is fully determined by the mode function $\f_k$ obeying \eq{me} and \eq{wc}. In particular, the vacuum energy density and pressure can be calculated from the momentum integrals involving $\f_k$ and $d\f_k/dt$ (see e.g. Ref. \onlinecite{ak2}). However, \eq{me} and \eq{wc} do not uniquely specify $\f_k$. Since \eq{me} is a second-order differential equation, it admits two linearly independent solutions. The general solution can therefore be written as a linear combination of these solutions, with two complex constants. The Wronskian condition \eq{wc} imposes only a single real constraint because it is identical to its own complex conjugate. Consequently, the choice of $\f_k$ retains a freedom corresponding to three arbitrary parameters (four real numbers from the two complex constants minus one real constraint).

This arbitrariness directly affects the determination of the vacuum state. Changing $\f_k$ (as defined in \eq{mex}) 
amounts to modifying the creation and annihilation operators and thus, according to  \eq{vac}, the vacuum state itself. As $\f_k$ satisfies the linear second-order differential equation \eq{me}, any such change can be expressed as a rank-2 linear map known as a Bogoliubov transformation, which linearly mixes the creation and annihilation operators of one basis with those of another. Therefore, the canonical quantization procedure alone does not uniquely determine the ground state; this is the well-known vacuum non-uniqueness of quantum fields in curved spacetime (see e.g. Ref. \onlinecite{bd}).

The ambiguity is resolved in flat spacetime by demanding that the vacuum state be invariant under Poincare symmetry (space and time translations, rotations, and boosts), which uniquely determines the mode function to be
\be\label{fsmf} 
\f_k=\fr{1}{\sqrt{2\o_k}}e^{-i\o_k t},\hs{10}a=1 \,\,(\textrm{flat space}).
\ee 
However a general cosmological background does not possess Poincare invariance; for example the time translation symmetry is explicitly broken in the metric \eq{frw}. Consequently, the condition of Poincare invariance cannot be used to uniquely define the vacuum state in curved spacetime.

A physically natural condition for defining the vacuum is to require the mode function $\f_k$ to behave like its flat-space counterpart \eq{fsmf} as $k\to\infty$. This is based on the expectation that the effects of cosmic expansion become negligible in the short-wavelength limit. From the two linearly independent solutions of \eq{me}, one can always choose a unique linear combination that satisfies this requirement. The corresponding vacuum state determined by this choice is known as the Bunch–Davies vacuum.
 
However, this specification is not generally valid at all times. For example, a state that is prescribed to be the Bunch–Davies vacuum during a radiation dominated era will evolve into a state that is no longer a Bunch–Davies vacuum during a subsequent matter domination. This is sometimes expressed by saying that, in general, it is impossible to maintain a purely negative-frequency solution at all times, where a negative-frequency solution is characterized by the time-dependence of the flat-space solution \eq{fsmf}. 

The non-uniqueness of the vacuum or  the fact that the vacuum state defined at one time need not remain a vacuum at a later time can be attributed to the non-adiabatic evolution of the mode function $\f_k$. From \eq{me}, the evolution of $\f_k$ is subject to two competing effects: both the $m^2$ and $k^2/a^2$ terms act as restoring forces that cause $\f_k$ to oscillate, while the term involving the Hubble parameter $H$ acts as a friction force that damps these oscillations. Significant damping is the primary source of non-adiabatic evolution. Therefore, the adiabatic approximation is valid provided
\be\label{adc}
\frac{k}{a} \gg H \qquad \text{or} \qquad m \gg H \qquad \text{at all times,}
\ee
so that the damping always remains  negligible. 

There are epochs in the early universe during which the Hubble parameter is much larger than the mass of the heaviest known elementary particle. For example, a typical assumed value of the Hubble parameter during inflation is $H_I=10^{13}\,$ GeV, whereas the mass of the heaviest known elementary particle, the top quark, is $173\,$ GeV. Consequently, the adiabatic approximation employed in Section  \ref{V} breaks down for modes satisfying $k/a< H_I$. This corresponds  to a broad range of physical momenta, for example $k_{\phys}<10^{13}\,$GeV, assuming the Planck mass sets the ultimate cutoff, $k_{\phys}<M_{pl}\,c^2\simeq 1.22\times 10^{19}\,$ GeV.

\section{More on  Regularization}

As mentioned above, the vacuum energy density in an expanding universe can be calculated from $\f_k$, so one may write $\r=\r[\f_k]$. Not surprisingly, in nearly all cases $\r[\f_k]$ diverges and therefore must be regularized, as before. 

In the standard cosmological picture, the universe comes into existence after the Big Bang and then undergoes inflation, radiation-dominated and matter-dominated eras, followed by the recent dark-energy-dominated accelerated expansion. There are no known analytic solutions for the mode function $\f_k$ valid over the entire history of the universe. Consequently, it is impossible to employ dimensional regularization to determine the cosmological evolution of the vacuum energy density.

There is an alternative, physically well-motivated method worth mentioning here, known as adiabatic regularization.\cite{ad} The main idea is that the vacuum energy density of an expanding universe is meaningful only when compared with that of flat spacetime or a slowly expanding universe; in other words, only the difference from a chosen reference state is physically significant. This is precisely the same reasoning used in the calculation of the Casimir effect.

To determine the energy density of the reference slowly expanding spacetime, one introduces the mode function in the adiabatic form
\be\label{admf} 
\f_k^{[ad]}=\fr{1}{\sqrt{2a^3\,\O_k}}\,e^{-i\int^t \O_k(t')dt'}.
\ee
This structure is derived from the flat-space mode function \eq{fsmf}, where $\O_k$ is the natural generalization of $\o_k$ in a time dependent background. The $a(t)$-dependence of the prefactor ensures that \eq{admf} identically satisfies the Wronskian condition \eq{wc}.  This mode is referred to as adiabatic because, in the limit where the expansion is switched off (i.e. $a(t)\to 1$), one has $\O_k\to\o_k$ and \eq{admf} reduces to the flat space mode function. 

Substituting \eqref{admf} into \eqref{me}, and treating the number of time derivatives acting on $a(t)$ and $\O_k$ as a perturbation parameter, one can obtain a series solution for $\O_k$ where the $n$'th order term contains exactly $n$ time derivatives (the $a(t)$ prefactor introduced in \eq{admf} also ensures that only even orders appear). For example, one can determine  $d\f^{[ad]}_k/dt$ from \eq{admf} as 
\be\label{ftd}
\fr{d\f^{[ad]}_k}{dt}=-\fr32 H \,\f_k^{[ad]} - \fr{1}{2\O_k}\fr{d\O_k}{dt}\,\f_k^{[ad]}-i\O_k\,\f_k^{[ad]}.
\ee
The first two terms contain time derivatives and are thus of a higher order compared to the last term, which contains no time derivatives. Accordingly, they are identified as being first and zeroth adiabatic orders, respectively. Taking an additional time derivative and substituting these expressions in \eq{me}, one finds that the zeroth-order solution for $\O_k$ is
\be
\O_k^{[0]}=\sqrt{\fr{k^2}{a^2}+m^2}. 
\ee
Note that the second term in \eqref{me} is already of second adiabatic order.

To regularize the vacuum energy density, or in fact any other expectation value in the theory, one subtracts the contribution obtained from \eq{admf}, calculated perturbatively to a sufficient order. Thus, one defines the regularized expression
\be\label{as} 
\r_{reg}=\r[\f_k]-\r[\f^{[ad]}_k],
\ee
where the series in $\r[\f^{[ad]}_k]$ should be kept up to the fourth adiabatic order to cancel the infinities. These are known as the adiabatic subtraction terms. 

While adiabatic regularization provides a physically meaningful method for removing divergences, it is important to recognize that discrepancies with other regularization techniques can arise. For example, although dimensional regularization yields \eq{r5} for the energy density of a massive scalar field in flat spacetime, adiabatic regularization using \eq{as} gives $\r_{reg}=0$.  A similar discrepancy occurs in the regularization of the scalar-field two-point function during inflation.\cite{par} Addressing these differences is an additional challenge associated with the cosmological constant problem.

\section{Conclusions}

The cosmological constant problem stands as one of the central, long-standing unsolved problems in modern physics. In this paper, we discuss the roots of the problem in elementary terms and highlight several conceptual difficulties associated with it. Although many proposals have been put forward, no satisfactory or generally accepted solution has yet been found. If dark energy is  a cosmological constant, which is observationally the most plausible possibility, there also arises the question of why its magnitude is comparable to the present mass density. This issue is sometimes referred to as the “new cosmological constant problem”.\cite{we2}

A satisfactory solution to the cosmological constant problem likely requires a major paradigm shift in physics, which would entail the quantization of gravity. What is commonly referred to as the vacuum energy density is, in fact, the expectation value of the energy density operator $\hat{\r}$ in the ground state $\left|0\right>$, namely  $\r=\left<0\right|\hat{\r}\left|0\right>$. However, it has been shown that there are order-one fluctuations around this mean value.\cite{ak3} Consequently, representing the quantum energy density  by its expectation value in a given state is not a good approximation. Hence, treating gravity classically while replacing the quantum vacuum energy density with its average value is questionable from the outset.

At present, there are two leading candidates for a quantum theory of gravity: string theory \cite{gsw} and loop quantum gravity.\cite{ash} Although their underlying assumptions differ significantly, both approaches are consistent with the fundamental principles of quantum mechanics.

On the other hand, the cosmological constant problem may point to issues of a fundamentally different nature, related to the specification of the initial state of the universe, an aspect on which quantum mechanics itself offers no direct guidance.\cite{ik1,ik2} As a result, even the discovery of a consistent quantum theory of gravity may not, by itself, resolve the cosmological constant problem. While this underscores the severity of the situation, it also suggests that the next major revolution in physics may lead to entirely unexpected surprises.

\end{document}